\documentclass[12pt,a4paper]{article}   
\usepackage{amsfonts}
\usepackage{color}
\usepackage{amssymb}
\usepackage{amsmath}
\usepackage{graphicx,color}
\usepackage{amsthm}
\usepackage[latin9]{inputenc}
\usepackage{enumerate}
\usepackage{bm}
\usepackage{fancyhdr}
\usepackage{pdfsync}
\newtheorem*{lemma*}{Lemma}

\numberwithin{equation}{section}
\date{}

\newcommand{\xes}{\textrm{\textbf{x}}}

\newcommand{\be}{\begin{equation}}\newcommand{\ee}{\end{equation}}
\newcommand{\bea}{\begin{eqnarray}}\newcommand{\eea}{\end{eqnarray}}
\newcommand{\nn}{\nonumber}
\newcommand{\pa}{\partial}
\newcommand{\Ref}[1]{(\ref{#1})}
\renewcommand{\ni}{\noindent}
\newcommand{\ep}{\varepsilon}

\begin{document}

\title{\bf Quantum vacuum interaction between two cosmic strings revisited}

\author{J. M. Mu$\tilde{\rm n}$oz-Casta$\tilde{\rm n}$eda$^1$\footnote{jose.munoz-castaneda@uni-leipzig.de}, and M. Bordag$^1$\footnote{bordag@itp.uni-leipzig.de}\\
\footnotesize{{\sl $^1$Institut f\"ur Theoretische Physik, Universit\"at Leipzig, Germany.}}}

\maketitle
\today
\begin{abstract}
    We reconsider the quantum vacuum interaction energy between two straight parallel cosmic strings. This problem was discussed several times in an approach treating both strings perturbatively and treating only one perturbatively. Here we point out that a simplifying assumption made in \cite{Bordag:1990if} can be justified and show  that, despite the global character of the background, the perturbative approach delivers a correct result. We consider the applicability of the scattering methods, developed in the past decade for the Casimir effect, for the cosmic string and find it not applicable. We calculate the scattering T-operator on one string. Finally, we consider the vacuum interaction of two strings when each carries a two dimensional delta function potential.
\end{abstract}

\section{Introduction}
%
Cosmic strings are static solutions of the Einstein equations which may have been created in the early universe. For one straight string the metric in conformal coordinates is a flat one with an angular deficit in the perpendicular plane making the space conical. As a result there are no local, but only global gravitational effects. For instance, there are tidal forces acting on a test particle \cite{line86-33-1833}, but there is no gravitational interaction between straight parallel strings \cite{lete87-4-75}.

In the background of cosmic strings quantum fluctuations of matter fields occur. Vacuum polarization for a single string, including the ultraviolet divergences and the back reaction were studied in quite a number of papers. Scattering in conical space was investigated in \cite{dese88-118-495} resulting in a simple expression for the scattering phase shift. The Green function in conical space time was considered in \cite{dowk77-10-115}. The corresponding energy-momentum tensor was calculated in \cite{hell86-34-1918,line87-35-1987} for scalar fields and in \cite{frol87-35-3779} for higher spins. For two or more parallel strings the vacuum polarization has a dependence on the separation and results in a Casimir effect. This was first considered in \cite{Bordag:1990if} and, independently, in \cite{galt95-58-3903} for scalar fields and later in \cite{alie97-55-3903} for the electromagnetic field. As a result, an attractive Casimir force comes to the zero classical force between the strings.
The calculations in \cite{galt95-58-3903,alie97-55-3903} were done   treating the mass densities $\lambda_i$ ($i=1,2)$ of two strings as a perturbation, i.e., in order $\lambda_1\lambda_2$.  In \cite{Bordag:1990if} the approach was used to take one string ($\lambda_1$) exact and only the second perturbatively. This is possible since the metric of one string is locally flat in conformal coordinates and from the string 'only' a reduced angular region follows. In this case the variables separate and simple integral representation can be used. The Casimir energy becomes proportional to $f(\lambda_1)\lambda_2$ with some function $f(\lambda_1)$, which is quite simple and regular (see eq. \Ref{5.36} below). For small $\lambda_1$ both results coincide.

In the present paper we reconsider the vacuum interaction for two parallel cosmic strings. After introducing necessary notations we reconsider the approach of \cite{Bordag:1990if}. We point out a calculational mistake and obtain the correct result. Further, we discuss a term which was dropped in \cite{Bordag:1990if} as 'free space' contribution. We show that this dropping was correct, although not well founded there. The problem is in the non trivial global properties of a conical background.
Further we consider the vacuum energy in terms of the trace of the logarithm of the Green function (in \cite{Bordag:1990if} the vacuum energy was calculated from the energy-momentum tensor). In the past decade powerful scattering approach (or 'TGTG' formula) was developed and applied to the Casimir effect for a wide variety of configurations. We calculate here the corresponding T-operator. Although the scattering phase shift for a single string is well known, so far there was no expression for the T-operator, at least in flat coordinates.

As known, the heat kernel expansion carries the essential information about the ultraviolet divergences of the vacuum energy. Again, for a single string, this expansion was studied in a number of papers beginning with \cite{dowk89-30-770} (see also \cite{furs94-11-1431}), basically making use of simple expressions for the Green functions. This approach was also extended to cosmic strings of finite thickness \cite{khus99-59-064017}. Here we consider the heat kernel coefficients for 2 strings as a special case of the general formulas in curved background \cite{vass03-388-279}. For the coefficients this is a nonperturbative information. We show that the coefficients relevant for for the divergences do not have separation dependent contributions. This confirms that the Casimir force is finite for all $\lambda_1$ and $\lambda_2$.

In general, the 'TGTG'-approach provides in many cases a representation  of the  Casimir force which is free of divergences in any stage of the calculation thus allowing for a tractable numerical evaluation. Regrettably, for 2 cosmic strings this approach does not work as we show in Section 5.
Further in that section we reconsider the perturbative approach and the approach with one string taken exactly. We are forced to introduce a regularization and a somehow delicate cancellation of divergences appears. Finally, we consider, in flat space, the vacuum polarization from two dimensional delta functions on parallel lines as a situation analogous  to two strings. The resulting Casimir force turns out to be zero when the Dirac delta strings are infinitely thin.

\ni Throughout this paper we use geometric units with $G=\hbar=c=1$.

\subsection{The single cosmic string. Notation}
Quantum mechanincs of free scalar particles in the static curved space time generated by a cosmic string was studied in the 80's and the 90's in \cite{dese88-118-495,thoof88-117-685,guim94-310-297} (see also refs \cite{hind99-58-477,cope85-255-201,vile85-5-263} for reviews on different models for cosmic strings). The problem reduces to studying the scattering of free quantum particles across a conical singularity. In the language of quantum physics in curved space-time this system is given by the free Hamiltonian of scalar particles on a cone. In any case the problem is finally reduced to a 2 dimensional problem in quantum mechanics.

The infinitesimal space-time element generated by a cosmic string is given by
\begin{equation}
ds^2=-dt^2+dl^2,
\end{equation}
where $dl^2$ the space element of the cosmic string. In global Cartesian flat coordinates $dl^2$  is given by
\begin{equation}
dl^2=e^{-4 V(x)}\left(dx_1^2+dx_2^2\right)+dx_3^2.
\end{equation}
This space element can also be written in global cylindrical coordinates $\{r,\theta\}$ with $r\in[0,\infty)$ and $\theta\in[0,2\pi]$,
\begin{equation}
dl^2=e^{-4 V(x)}\left(dr^2+r^2d\theta^2\right)+dx_3^2.
\end{equation}
As is written in \cite{dese88-118-495}, the potential $V(x)$ generated by a cosmic string placed at the origin is given by
\begin{equation}
V(x)= 2\lambda \log(r).\label{1.4}
\end{equation}
The physical meaning of the constant $\lambda$ is the dimensionless mass density of the cosmic string. Restoring dimensions it would be   $\mu G/c^2$. Defining the angular deficit $\alpha$ by
\begin{equation}
\alpha=4\lambda+1,\label{1.5}
\end{equation}
the space element $dl^2$ reads
\begin{equation}
dl^2=\frac{1}{r^{2(\alpha-1)}}\left(dr^2+r^2d\theta^2\right)+dx_3^2.
\end{equation}
This formula is exactly the same as formula (2.1) in reference \cite{dese88-118-495}.
By the transformation
\begin{equation}
\rho=\alpha r^\alpha,\quad \psi=\alpha\theta \, \Rightarrow\quad \rho\in[0,\infty),\,\,\,\psi\in[0,2\pi\alpha]
\end{equation}
we switch to conformal coordinates. These have the property that the metric is locally flat.
In these coordinates points with $\psi=0$ and $\psi=2\pi\alpha$ are identified where $\alpha$ is the angular deficit \Ref{1.5}. The space element corresponding to the cosmic string is given by the expression
\begin{equation}
dl^2=d\rho^2+\rho^2d\psi^2+dx_3^2.
\end{equation}
From now on we will denote a space-time point in flat coordinates using Latin letters, e.g.  $x$, and a space-time point written in conformal coordinates using Greek letters, e.g. $\xi$. So for example $d^4x=dx^0 dx_1dx_2dx_3 =rdx^0 drd\theta dx_3$ and $d^4\xi=\rho dx^0 d\rho d\psi dx_3$. The volume element of the space-time generated by a cosmic string is given by:
\begin{equation}
d\mu_g=e^{-4V}d^4x=r^{-8\lambda}d^4x=r^{-2(\alpha-1)}d^4x=d^4\xi.
\end{equation}
Similarly we will denote by Greek characters a space-time point in conformal coordinates and by Latin characters a space-time point in flat coordinates.:
\begin{equation}
x=(x^0,x^1,x^2,x^3)=(x^0,r,\theta,x^3);\quad \xi=(x^0,\rho,\psi,x^3).
\end{equation}
The same notation with bold characters would be used to denote the pure space points. In addition we will denote points in the perpendicular planes to the string with bold characters, because it is a pure space point, and a subindex $\perp$. Therefore if the cosmic string lies on the $x_3$ axis  we have
\begin{eqnarray*}
{\bf x}=(x^1,x^2,x^3)=(r,\theta,x^3)& , & {\boldsymbol \xi}=(\rho,\psi,x^3),\\
{\bf x}_\perp=(x^1,x^2)=(r,\theta )& , & {\boldsymbol \xi_\perp}=(\rho,\psi).
\end{eqnarray*}
The   dynamics of a free massive scalar field in the classical background of a cosmic string is governed  by the action
\begin{equation}
  S(\phi)=\frac{1}{2}\int dx^4 r^{-8\lambda}\phi\left[-\partial_0^2+\partial_3^2+r^{8\lambda}\left(\partial_1^2+\partial_2^2\right)
  -m^2\right]\phi,\label{sca-actflat}
\end{equation}
written in flat coordinates. Using the conformal coordinates the action acquires a much simpler form,
\begin{equation}
  S(\phi)=\frac{1}{2}\int d\xi^4 \phi\left(-\partial_0^2+\partial_3^2+\frac{1}{\rho}\partial_\rho \rho\partial_\rho +
  \frac{1}{\rho^{2}}\partial_\psi^2-m^2\right)\phi  .  \label{sca-actcon}
\end{equation}
Formally the action (\ref{sca-actcon}) looks like the action for a free scalar field in the 4-dimensional Minkowski space-time. Nevertheless is not the same action because one must take into account that the angular periodicity is in the interval $\psi\in[0,2\pi\alpha]$ and therefore the space-time is conical. For instance, there is no translational invariance in the direction perpendicular to the string.

The Green function that arises from the action functional (\ref{sca-actcon}) can be found in many places, see for example refs. \cite{dese88-118-495,thoof88-117-685,guim94-310-297,Bordag:1990if}.
First, we make a Fourier transform in the translational invariant directions $\mu=0,3$,
\be\label{G1}
    G(\xi,\eta)=\int\frac{dk_0dk_3}{(2\pi)^2}e^{ik_s(\xi^s-\eta^s)}G_\gamma({\boldsymbol \xi_\perp},{\boldsymbol \eta_\perp})
\ee
with $s=0,3$ and $\gamma=\sqrt{-k_0^2+k_3^2+m^2}$.
Following \cite{Bordag:1990if}, for the radial dependence we use an  expansion in terms of modified Bessel functions with imaginary index, the known Kontorovich-Lebedev transform \cite{kont38-8-1192,sneddon},
\begin{equation}
G_\gamma({\boldsymbol \xi_\perp},{\boldsymbol \eta_\perp})=\int_{-\infty}^\infty\frac{d\mu}{\pi^2}\ e^{\pi\mu}\,K_{i\mu}(\gamma \rho_\xi) K_{i\mu}(\gamma \rho_\eta)G_\mu(\psi_\xi-\psi_\eta).     \label{green-1cs}
\end{equation}
We mention the equation
\be\label{Kimu} \left(\rho\pa_\rho \rho \pa_\rho+\rho^2\right)K_{i\mu}(\rho)=-\mu^2K_{i\mu}(\rho)
\ee
and the orthogonality relation
\be\label{orth}
    \int_{-\infty}^\infty \frac{d\mu}{\pi^2} \mu e^{\pi \mu}K_{i\mu}(\rho)K_{i\mu}(\rho')
    = \rho\,\delta(\rho-\rho').
\ee
The remaining angular Green function carries the information about the cosmic string and it satisfies the equation
\begin{equation}
\left(-\partial_\psi^2+\mu^2\right)G_\mu(\psi-\psi')=\delta(\psi-\psi')
\end{equation}
and it must be periodic in $\psi$ in the interval $[0,2\pi\alpha]$.
This equation can be easily solved by the expansion
\begin{equation}
    G_\mu(\Delta\psi)=
    \frac{1}{2\pi\alpha}\sum_{n=-\infty}^{\infty}
    \frac{\exp\left(i\,n\Delta\psi/\alpha\right)}{\mu^2+ n^2/\alpha^2} .  \label{green-ang-series}
\end{equation}
This series can be summed and with the restriction $|\Delta\psi|\leq 2\pi\alpha$  the angular Green function becomes
\be
    G_\mu(\Delta\psi)=
    \frac{1}{2 \mu }
    \left(\frac{\cosh(\mu\Delta\psi)e^{-\alpha\pi\mu}}{\sinh(\alpha\pi\mu)}
            +e^{-\mu|\Delta\psi|}\right).   \label{green-angular}
\ee

\subsection{The system of two cosmic strings and the approach of \cite{Bordag:1990if}}

Suppose now that we have a space-time defined by the presence of two parallel cosmic strings with mass densities $\lambda_1$ and $\lambda_2$. The spatial origin and axes are chosen in such a way that one  string is placed on the $x_3$ axis and the positions of the strings are given in rectangular flat coordinates by
\begin{equation}
{\bf s}_1=(0,0,0),\quad {\bf s}_2=(b,0,0).
\end{equation}
The spatial metric tensor in cylindrical coordinates is then given by
\begin{equation}
dl^2=e^{-4(V_1+V_2)}\left(dr^2+r^2d\theta^2 \right),\quad V_i=2\lambda_i \log(r_i),\label{1.21}
\end{equation}
where $r_i=|{\bf x}-{\bf s}_i|$ and $i=1,2$. Again the spatial sections orthogonal to both strings are conformally flat. The action for a scalar field propagating in this background is formally the same as given in equation (\ref{sca-actflat}) but replacing $V$ by $V_1+V_2$.

Since the scattering problem of scalar quantum particles in the background of one cosmic string was exactly solved in references \cite{dese88-118-495,thoof88-117-685} the approach taken  in reference \cite{Bordag:1990if} to compute the quantum vacuum interaction between two cosmic strings is to account for one of the strings exactly (all orders in $\lambda_1$ are taken into account) and for the other as a perturbation potential over the background of the first string (only first order in $\lambda_2$ contributions are taken into account). Under this assumption the action that governs the dynamics of a scalar field in the background of two cosmic strings has the structure
\begin{equation}
S(\phi)=S_0(\phi;\,\lambda_1)+S_{int}(\phi;\,\lambda_1,\lambda_2,b),
\end{equation}
being $S_0(\phi;\,\lambda_1)$ the action (\ref{sca-actcon}) written in conformal coordinates associated to the non-perturbative string. $S_{int}$ is the interaction term of the perturbative string with the classical background of the non-perturbative string written in conformal coordinates associated to the first string,
\begin{equation}
S_{int}=-2\lambda_2\int d\xi^4\phi(\xi)\log\left[ \left(\frac{\rho}{\beta}\right)^{\alpha_1}+1-2\cos\left(\psi/\alpha_1\right)\left(\frac{\rho}{\beta}\right)^{1/\alpha_1} \right](\partial^2_{\xi^3}-\partial^2_{\xi^0}+m^2)\phi(\xi),\label{sint}
\end{equation}
being $\beta$ the conformal distance between both strings,
\begin{equation}
\beta=b^{\alpha_1}/\alpha_1.
\end{equation}
In the approach taken  in reference \cite{Bordag:1990if}, the 1st cosmic string plays the role of the standard vacuum with which the quantum vacuum interaction energy is measured. In addition, the 2nd string plays the role of a perturbation (potential) over the reference vacuum defined by the first string. Under this assumptions we can write down the Born series for the Green function of two cosmic strings in terms of the Green function for the 1st string (it will play the role of $G_\omega^{(0)}$) and the interaction operator that determines $S_{int}$,
\footnote{From now on we will denote the Green function for the 1st cosmic string with the super-index $1s$, and similar for the second string.}
\begin{equation}
{\cal G}_\omega^{2s}={\cal G}_\omega^{1s}+{\cal G}_\omega^{1s}\cdot \widehat{{\cal L}}_{int}\cdot {\cal G}_\omega^{1s}+{\cal O}(\lambda_2^2).
\end{equation}
From equation (\ref{sint}) we can write the operator $\widehat{{\cal L}}_{int}$ in conformal coordinates associated to the first string as
\begin{equation}
\widehat{{\cal L}}_{int}=-2\lambda_2\log\left[ \left(\frac{\rho}{\beta}\right)^{\alpha_1}+1-2\cos\left(\psi/\alpha_1\right)\left(\frac{\rho}{\beta}\right)^{1/\alpha_1} \right](\partial^2_{\xi^3}-\partial^2_{\xi^0}+m^2). \label{lint}
\end{equation}
The vacuum energy energy density per unit length $E_0$ for the perturbation that arises from the second string in the background of the first string is given by very standard formulas in terms of the expectation value of the energy momentum tensor,
\begin{equation}
E_0=\int d\xi_1d\xi_2<T_{00}>=\left. \frac{1}{i}\int d\xi_1d\xi_2\left(-\partial_{\xi^0}^2+{1\over 2}(\partial_{\xi_\mu}\partial^{\xi_\mu}+m^2)\right)G^{2s}_\omega(\xi,\eta)\right\vert_{\eta=\xi}.
\end{equation}
When plugging the Born series expansion in terms of ${\cal G}_\omega^{1s}$  into the expression for the vacuum energy given above, one gets the expansion
\begin{eqnarray}
E_0&=&\left. \frac{1}{i}\int d\xi_1d\xi_2\left(-\partial_{\xi^0}^2+{1\over 2}(\partial_{\xi_\mu}\partial^{\xi_\mu}+m^2)\right)\left( \int d^4\xi' G^{1s}(\xi,\xi')\widehat{{\cal L}}_{int}(\xi')G^{1s}(\xi',\eta)\right)\right\vert_{\eta=\xi}\nonumber\\
&&+\left. \frac{1}{i}\int d\xi_1d\xi_2\left(-\partial_{\xi^0}^2+{1\over 2}(\partial_{\xi_\mu}\partial^{\xi_\mu}+m^2)\right)\left( G^{1s}(\xi,\eta)\right)\right\vert_{\eta=\xi}+{\cal O}(\lambda_2^2).\label{totvacener}
\end{eqnarray}
The two contributions up to 1$^{st}$ order in $\lambda_2$ to the total vacuum energy (\ref{totvacener}) have the following interpretation:
\begin{itemize}
\item The first term in (\ref{totvacener}) represents the quantum vacuum interaction energy between both strings up to first order in $\lambda_2$.
\item The second term in (\ref{totvacener}) represents the vacuum energy of the first string because there is no dependence on the mass density of the second string $\lambda_2$.
\end{itemize}
Therefore the quantum vacuum interaction energy density per unit length between both cosmic strings up to first order in the mass density $\lambda_2$ of the second string is given by
\begin{equation}
E_{int}=\left. \frac{1}{i}\int d\xi_1d\xi_2\left(-\partial_{\xi^0}^2+{1\over 2}(\partial_{\xi_\mu}\partial^{\xi_\mu}+m^2)\right)\left( \int d^4\xi' G^{1s}(\xi,\xi')\widehat{{\cal L}}_{int}(\xi')G^{1s}(\xi',\eta)\right)\right\vert_{\eta=\xi}.\label{intener}
\end{equation}
Up to this point all formulas are very standard and can be found for example in references like \cite{BKMM,milton01,full89b}. The idea of reference \cite{Bordag:1990if} is to use the representation of the Green function for a single cosmic string given by (\ref{green-1cs}) in the calculation of the quantum vacuum interaction energy density per unit length given by formula (\ref{intener}). In the procedure, the representation of the angular Green function given in equation (\ref{green-angular}) is used.

In the course of the calculations, the second term in the parentheses in eq. \Ref{green-angular} was identified as the contribution from the free space and it was dropped. The vacuum energy resulting from this term is potentially ultraviolet divergent since this contribution has no exponential decrease for large $\mu$ in the coincidence limit $\Delta\psi=0$ which must be taken in the vacuum energy in \Ref{totvacener} or \Ref{intener}. But this term does not depend on the angular deficit $\alpha$ of the first string. Thus it corresponds to the vacuum energy resulting   from the second string alone and would not contribute to the Casimir force and can be dropped. In this way, in \cite{Bordag:1990if} the ultraviolet divergences were removed. However,  this argumentation is not correct since such dependence is hidden in the conformal coordinates and it comes back if doing the transformation back to flat coordinates. Therefor the result of \cite{Bordag:1990if} must be questioned.
It must be mentioned that the subtraction of the 'free space contribution' to the Casimir energy is in most cases justified as not contributing to the Casimir force. In this sense, the cosmic string is exceptional due to the global character of the conic space time.

However, from the calculation done in this paper it turns out that the 'free space contribution', dropped without further justification in \cite{Bordag:1990if}, in fact evaluates to zero, justifying the procedure used in \cite{Bordag:1990if}.
Regretably, in \cite{Bordag:1990if} there is a calculational mistake invalidating the result obtained. This mistake can be identified as a sign error in the equation between eqs.(26) and (27) in \cite{Bordag:1990if} and after correction one would obtain the same result as below in eq. \Ref{5.36}.

\section{The $TGTG$ setup for curved static space-times.}
In general, the scattering or 'TGTG'-setup for the vacuum energy starts from the equation
\be
\left(-\pa_{x_0}^2+\Delta+V(x)\right)G^{(V)}(x,x')=\delta^{(4)}(x-x')
\label{2.0} \ee
for a background potential $V(x)$ (see for example, chap.10 in \cite{BKMM} or \cite{kenn08-78-014103}). For 2 cosmic strings such structure can be achieved in flat coordinates. First, note the equation
\be
\left( -\partial_0^2+\pa_z^2+\partial_r^2+\frac{1}{r}\partial_r+\frac{1}{r^2}\partial_\theta^2\right) G^{(0)}(x,x')
\equiv\left( -\partial_0^2+\Delta_{flat}\right)G^{(0)}(x,x')=\delta^{(4)}(x-x')
\label{2.1}\ee
for the free Green function in flat space written in cylindrical coordinates. In these notation, the equation
\be
\left( \left(-\partial_0^2+\pa_z^2\right)\,e^{-4(V_1+V_2)}
    +\partial_r^2+\frac{1}{r}\partial_r+\frac{1}{r^2}\partial_\theta^2\right)G^{(V)}(x,x')
    =\delta^{(4)}(x-x')
\label{2.2}\ee
for the Green function in the background of cosmic strings with a potential given by eq. \Ref{1.4} can be rewritten in the form
\be
 \left(-\partial_0^2+\Delta_{flat}
    +\left(\,e^{-4(V_1+V_2)}-1\right)
   \left(-\partial_0^2+\pa_z^2\right)
   \right)G^{(V)}(x,x')
    =\delta^{(4)}(x-x'),
\label{2.3}\ee
which is eq.\Ref{2.0} with the substitution
\be V(x)\to \left(\,e^{-4(V_1+V_2)}-1\right)
   \left(-\partial_0^2+\pa_z^2\right).
\label{2.4}\ee
Now, the essence of the 'TGTG'formula is to rewrite the vacuum energy, given by (see, for example, eq.(10.22) in \cite{BKMM})
\be E_0=-\frac12 Tr\ln G^{(V)},
\label{2.5}\ee
in the form
\be E_0=\frac12 Tr\left(1-G^{(0)}T_A G^{(0)} T_B\right)+(\mbox{separation independent contributions})
\label{2.6}\ee
in case the potential splits accordingly to
\be V=V_A+V_B
\label{2.7}\ee
into parts corresponding to the objects whose vacuum interaction we are interested in.
The merit of \Ref{2.6} is that all ultraviolet divergent contributions are dropped together with the separation independent ones. Thus the remaining trace has no divergences allowing for a tractable numerical evaluation.

As for the cosmic strings, for two of them, the potential is given by \Ref{2.4} with \Ref{1.21}. While the derivatives can be handled by Fourier transform, the structure is not the sum of two potentials like in eq. \Ref{2.7}, but rather a product. Therefore the potential \Ref{2.4} is not additive, rather a product. As a consequence, no formula like \Ref{2.6} can be derived (except for in lowest perturbative order).

Despite this regrettable observation, it is of interest to study the $T$-operator for a single string, which will be done in the remainder of this section. After having defined by \Ref{2.1} the free Green function in flat coordinates, we define by \Ref{2.2} the Green function for one string with $V$ being given by \Ref{1.4}. In both cases, the 4-dimensional   delta function is given by
\be
\delta^{(4)}(x-x')= \delta(x^0-x'^0)\delta(x^3-x'^3) \frac{1}{r}\delta(r-r') \delta(\theta-\theta').
\ee
Since for the flat space-time and for the cosmic string space-time the time coordinate $x^0$ and the space coordinate $x^3$ are flat we can write
\be
G^{(0)}(x,x')=\int_{-\infty}^\infty\frac{d\omega}{2\pi}e^{i\omega(x^0-x'^0)}\int_{-\infty}^\infty\frac{dk_3}{2\pi}e^{ik_3(x^3-x'^3)}G^{(0)}_\gamma({\bf x}_\perp ,{\bf x}'_\perp),
\ee
\be
G(x,x')=\int_{-\infty}^\infty\frac{d\omega}{2\pi}e^{i\omega(x^0-x'^0)}\int_{-\infty}^\infty\frac{dk_3}{2\pi}e^{ik_3(x^3-x'^3)}G_\gamma({\bf x}_\perp ,{\bf x}'_\perp),
\ee
where $\gamma=\sqrt{-\omega^2+k_3^2}$. The equations for the orthogonal Green functions $G^{(0)}_\gamma({\bf x}_\perp ,{\bf x}'_\perp)$ and $G_\gamma({\bf x}_\perp ,{\bf x}'_\perp)$ are now given by
\be
\left(-\gamma^2+\Delta_{flat}^\perp \right)G^{(0)}_\gamma({\bf x}_\perp ,{\bf x}'_\perp)=\frac{1}{r}\delta(r-r') \delta(\theta-\theta') \label{eq-ggamma0},
\ee
\be
\left(-\gamma^2+\Delta_{flat}^\perp -(e^{-4V}-1)\gamma^2\right)G_\gamma({\bf x}_\perp ,{\bf x}'_\perp)=\frac{1}{r}\delta(r-r') \delta(\theta-\theta'),
\label{eq-ggamma}
\ee
being $\Delta_{flat}^\perp =\partial_r^2+\frac{1}{r}\partial_r+\frac{1}{r^2}\partial_\theta^2$ the Laplace operator in the 2 dimensional orthogonal sections to the cosmic string. At this point it is useful to introduce the  notations
\be
\widehat{\mathcal{O}}^{(0)}_\gamma({\bf x}_\perp)\equiv-\gamma^2+\Delta_{flat}^\perp,
\ee
\be
\widehat{\mathcal{O}}_\gamma({\bf x}_\perp)\equiv-\gamma^2+\Delta_{flat}^\perp -(e^{-4V}-1)\gamma^2= \widehat{\mathcal{O}}^{(0)}_\gamma({\bf x}_\perp)-(e^{-4V}-1)\gamma^2.
\ee
Following the section 4.1 in reference \cite{bord12-45-374012} we now introduce the $T$-operator that describes the propagation of free scalar particles in the classical background of a single cosmic string. The formal definition of the $T$-operator in the orthogonal sections to the cosmic string is given by writing $G_\gamma$ in terms of $G_\gamma^{(0)}$ and the $T$-operator that satisfies the relation
\be
G_\gamma({\bf x}_\perp ,{\bf x}'_\perp)\equiv G^{(0)}_\gamma({\bf x}_\perp ,{\bf x}'_\perp)-\int d{\bf z}_\perp d{\bf z}'_\perp G^{(0)}_\gamma({\bf x}_\perp ,{\bf z}'_\perp)T_\gamma({\bf z}_\perp ,{\bf z}'_\perp)G^{(0)}_\gamma({\bf z}'_\perp ,{\bf x}'_\perp).
\ee
If we take into account that $\widehat{\mathcal{O}}^{(0)}_\gamma({\bf x}_\perp)\cdot G^{(0)}_\gamma({\bf x}_\perp ,{\bf x}'_\perp)=\delta(r-r')\delta(\theta-\theta')/ r$ then it is straight forward to obtain the following expression for the kernel of the $T$-operator in the flat othogonal sections to the string,
\be
T_\gamma({\bf x}_\perp ,{\bf y}_\perp)=-\widehat{\mathcal{O}}^{(0)}_\gamma({\bf z}_\perp)\widehat{\mathcal{O}}^{(0)}_\gamma({\bf z}'_\perp)\left(G_\gamma({\bf z}_\perp ,{\bf z}'_\perp)-G^{(0)}_\gamma({\bf z}_\perp ,{\bf z}'_\perp)\right).\label{gen-tgamma}
\ee
The expression for $G^{(0)}_\gamma({\bf z}_\perp ,{\bf z}'_\perp)$ in cylindrical coordinates is given by\footnote{$G^{(0)}_\gamma({\bf z}_\perp ,{\bf z}'_\perp)$ can be directly obtained from equations \Ref{green-1cs} and \Ref{green-angular} making the angular deficit $\alpha=1$.}
\be
G^{(0)}_\gamma({\bf z}_\perp ,{\bf z}'_\perp)=\int_{-\infty}^\infty\frac{d\mu}{\pi^2}\mu e^{\pi\mu}\,K_{i\mu}(\gamma r) K_{i\mu}(\gamma r')G^{(0)}_\mu(\theta-\theta')
\ee with
\be
G^{(0)}_\mu(\theta-\theta')=\frac{1}{2 \mu }
    \left(\frac{\cosh(\mu(\theta-\theta'))e^{-\pi\mu}}{\sinh(\pi\mu)}
            +e^{-\mu|\theta-\theta'|}\right).
\ee
Similarly, using equations \Ref{green-1cs} and \Ref{green-angular} written in cylindrical coordinates instead of conformal coordinates, we obtain for $G^{(0)}_\gamma({\bf z}_\perp ,{\bf z}'_\perp)$ the expression
\be
G_\gamma({\bf z}_\perp ,{\bf z}'_\perp)=\int_{-\infty}^\infty\frac{d\mu}{\pi^2}\mu e^{\pi\mu}\,K_{i\mu}(\gamma\alpha r^\alpha) K_{i\mu}(\gamma\alpha r'^\alpha)G_\mu(\theta-\theta') \label{ggamma-cy}
\ee with
\be
G_\mu(\theta-\theta')=\frac{1}{2 \mu }
    \left(\frac{\cosh(\mu\alpha(\theta-\theta'))e^{-\alpha\pi\mu}}{\sinh(\alpha\pi\mu)}
            +e^{-\mu\alpha|\theta-\theta'|}\right).\label{angggamma-cy}
\ee
From expression (\ref{gen-tgamma}) and using equations (\ref{eq-ggamma0}) and (\ref{eq-ggamma}) we can obtain an explicit expression for the kernel of the $T$-operator\footnote{Note that $\widehat{\mathcal{O}}^{(0)}_\gamma({\bf z}_\perp)G_\gamma({\bf z}_\perp ,{\bf z}'_\perp)= \delta(r-r')\delta(\theta-\theta')/ r+(e^{4V(r)}-1)\gamma^2G_\gamma({\bf z}_\perp ,{\bf z}'_\perp)$} in terms of  $G_\gamma({\bf z}_\perp ,{\bf z}'_\perp)$,
\be
T_\gamma({\bf z}_\perp ,{\bf z}'_\perp)=-\gamma^2\frac{e^{-4V}-1}{r}\delta(r-r')\delta(\theta-\theta')-\gamma^4(e^{-4 V(r)}-1)(e^{-4 V(r')}-1)G_\gamma({\bf z}_\perp ,{\bf z}'_\perp). \label{exp-Tkernel}
\ee

Now we can compute the radial components of operator-$T$. First of all we make a decomposition of $T_\gamma({\bf z}_\perp ,{\bf z}'_\perp)$ in the angular basis\footnote{Take into account that the cosmic string is a system with cylindrical symmetry.}
\begin{equation}
T_\gamma({\bf z}_\perp ,{\bf z}'_\perp)=\sum_{L=-\infty}^{\infty}\, e^{i L (\theta-\theta')}T_\gamma^{(L)}(r,r').\label{2.17}
\end{equation}
Hence, accounting for  integral form of the Kronecker delta,
\begin{equation}
\delta_{LM}=\frac{1}{2\pi}\int_0^{2\pi}d\theta \,e^{i\theta (L-M)},
\end{equation}
the radial components can be  written in terms of the complete kernel as
\begin{equation}
T_\gamma^{(L)}(r,r')=\int_0^{2\pi}\frac{d\theta}{2\pi}\, e^{-i L\theta}\left.T_\gamma({\bf z}_\perp ,{\bf z}'_\perp)\right\vert_{\theta'=0}.
\end{equation}
If we assume that the integration in $\theta$ commutes with the integration in $\mu$ coming from the expression \Ref{exp-Tkernel} for the kernel of operator ${\cal T}_\gamma$ then we can express the radial components of $T_\gamma^{(L)}(r,r')$ as
\begin{eqnarray}
T_\gamma^{(L)}(r,r')&=&-\gamma^2\frac{e^{-4V}-1}{2\pi r}\delta(r-r')-\gamma^4(e^{-4 V(r)}-1)(e^{-4 V(r')}-1)\nonumber\\
&&\times \int_{-\infty}^\infty\frac{d\mu}{\pi^2} \mu \, e^{\pi\mu}\,K_{i\mu}(\gamma\alpha r^\alpha) K_{i\mu}(\gamma\alpha r'^\alpha)\int_0^{2\pi}\frac{d\theta}{2\pi}e^{-iL\theta} G_\mu(\theta).
\end{eqnarray}
The integration over $\theta$ can easily be done  resulting in
\begin{equation}
\int_0^{2\pi}\frac{d\theta}{2\pi}e^{-iL\theta} G_\mu(\theta)= \frac{\alpha }{\alpha ^2 \mu ^2+L^2}.
\end{equation}
Hence the radial components for the kernel of operator ${\cal T}_\gamma$ are given by
\begin{eqnarray}
T_\gamma^{(L)}(r,r')&=&-\gamma^2\frac{e^{-4V}-1}{2\pi r}\delta(r-r')-\gamma^4(e^{-4 V(r)}-1)(e^{-4 V(r')}-1)\nonumber\\
&&\times\int_{-\infty}^\infty\frac{d\mu}{\pi^2}\, e^{\pi\mu}\,K_{i\mu}(\gamma\alpha r^\alpha) K_{i\mu}(\gamma\alpha r'^\alpha)\frac{\alpha\mu }{\alpha ^2 \mu ^2+L^2}.
\end{eqnarray}
Using reference \cite{grad94} we obtain
\be
\int_{-\infty}^\infty\frac{d\mu}{\pi^2} e^{\pi\mu}\,K_{i\mu}(\gamma\alpha r^\alpha) K_{i\mu}(\gamma\alpha r'^\alpha)\frac{\alpha\mu }{\alpha ^2 \mu ^2+L^2}=\frac{1}{\alpha} I_{\frac{L}{\alpha}}(\gamma\alpha r_<^\alpha) K_{\frac{L}{\alpha}}(\gamma\alpha r_>^\alpha),
\ee
where $r_>={\rm max}(r,r')$ and $r_<={\rm min}(r,r')$. Therefore finally using that $-4V=\log(r^{-2 (\alpha -1)})=\log(r^{-8\lambda})$ we obtain
\be
T_\gamma^{(L)}(r,r')=-(r^{-2(\alpha-1)}-1)\left(\gamma^2\frac{\delta(r-r')}{2\pi r}+\frac{\gamma^4}{\alpha}(r'^{-2(\alpha-1)}-1) I_{\frac{L}{\alpha}}(\gamma\alpha r_<^\alpha) K_{\frac{L}{\alpha}}(\gamma\alpha r_>^\alpha)\right)
\ee
which, being inserted into \Ref{2.17}, gives the $T$-operator.

%
\section{Preparation for a heat kernel approach I: general formulas}

\subsection{General geometrical formulas}
In order to approach the problem of the vacuum interaction between two cosmic strings we write down general formulas relating geometrical tensors of two Riemannian manifolds conformally related. In particular we are interested on the calculation of the geometrical objects associated to a conformally flat manifold. Let $M$ be a smooth manifold with two different Riemannian structures given by metric tensors $g$ and $\widetilde g$. Let us assume that $g$ and $\widetilde g$ are conformally equivalent, i. e.,
\begin{equation}
 \widetilde{g}=e^{2\epsilon F} g,
\end{equation}
where $F$ is a scalar function over $M$ and $\epsilon$ a positive arbitrary parameter. The two basic formulas that relate geometrical objects associated to $g$ with the ones associated to $\widetilde g$ are
\begin{enumerate}[i.]
\item Christoffel symbols
\begin{equation}
\widetilde\Gamma^i_{jk}=\Gamma^i_{jk}+\epsilon\left[\delta^i_j\partial_k F+\delta^i_k\partial_j F-g^{il}g_{jk}\partial_l F\right],
\end{equation}
\item Riemann curvature tensor
\begin{eqnarray}
\widetilde R^i_{jkl}&=&R^i_{jkl}+\epsilon\left[\delta^i_l\nabla_j\nabla_k-\delta^i_k\nabla_j\nabla_l +g_{jk} \nabla^i\nabla_l- g_{jl}\nabla^i\nabla_k\right]F\nonumber\\
&+&\epsilon^2\left[-\delta^i_l\nabla_j F\nabla_k F+g_{jl}\nabla^i F\nabla_k F+\delta^i_l g{jk}\nabla_n F\nabla^n F\right.\nonumber\\
&+&\left.\delta^i_k\nabla_l F\nabla_j F-g_{jk}\nabla_l F\nabla^i F-\delta^i_k g_{jl}\nabla_n F\nabla^n F\right],
\end{eqnarray}
\item Ricci tensor
\begin{eqnarray}
\widetilde R_{jk}&=&R_{jk}-\epsilon\left[(D-2)\nabla_j\nabla_k+g_{jk} \Delta\right]F\nonumber\\
&+&\epsilon^2\left[\nabla_j F\nabla_k F-g_{jk}\nabla_n F\nabla^n  F\right],
\end{eqnarray}
\item Ricci scalar
\begin{equation}
\widetilde R=e^{-2\epsilon F}\left[R- 2\epsilon (D-1)\Delta F -\epsilon^2(D-1)(D-2)\nabla_n F\nabla^n  F\right].
\end{equation}
\end{enumerate}
All these formulas have been taken from reference \cite{kirs01b} (appendix B).

\subsection{Heat kernel coefficients}
Due to the factorisation properties of the heat kernel (see for example reference \cite{kirs01b} for a review of results) the heat kernel coefficients for a Laplace-type operator defined over the spacetime of two parallel cosmic strings is fully determined by the heat kernel of the operator restricted to the orthogonal surfaces to both strings. 
The divergent part of the vacuum energy,
\be E_0^{\rm div}
    =\frac{a_2}{4\pi^{3/2}}\frac{1}{\delta^2}+\frac{a_4}{16\pi^2}\ln\delta,
\label{4.32}\ee
following eq. (4.32) for $m=0$ in \cite{BKMM} and adopting to our notations by substituting $a_{n}\to a_{2n}$ and having $\delta$ as regularization parameter, is determined by the heat kernel coefficients $a_2$ and $a_4$ of the Laplace-Beltrami operator defined over the orthogonal surfaces to both strings.

The review \cite{vass03-388-279} contains general formulas to compute the heat kernel coefficients of a large number of Laplace-type operators defined over different types of Riemannian manifolds. In our case we focus or attention on Laplace-type operators that govern the scalar quantum fluctuations over a conic 2D manifold without gauge fields and classical background classical fields. We follow notation and conventions used in   \cite{vass03-388-279}. We're only interested in the $a_2$ and $a_4$ coefficients and mention the following specifications,
\begin{itemize}
\item Because we are dealing with scalar quantum fluctuations the spin connection that defines or quantum field is trivial: $\omega=0\Rightarrow\Omega=0$,
\item Since there are no background gauge fields $\Rightarrow$ there is no principal fiber bundle structure. Hence the only geometrical coupling is the conformal coupling: $E=-\xi R$,
\item Since the quantum scalar field has no flavour degrees of freedom the Laplace-type operator that governs the dynamics of the one particle states is defined over a trivial line bundle. Due to the trivial line bundle structure all algebraic traces become trivial  in the formulas in \cite{vass03-388-279}.
\end{itemize}
With these assumptions the formulas we must use for the $a_2$ and $a_4$ coefficients are given by
\begin{equation}
(4\pi)^{D/2}a_2=\frac{1}{6}\int_M d^Dx\sqrt{g}\,(1-\xi)R
\end{equation}
and
\begin{eqnarray}
(4\pi)^{D/2}a_4&=&\frac{1}{360}\int_M d^Dx\sqrt{g}\left[12 (1-5\xi)\Delta R+(5+180\xi^2-60\xi)R^2\right.\nonumber\\
&&~~~~~~~~~~~~~~~~~~~~~~~~~-\left. 2R_{ij}R^{ij}+2R^{ijkl}R_{ijkl}\right]\label{gena4}.
\end{eqnarray}
These formulas will allow us to compute the heat kernel coefficients of the Laplace-type operator that governs the dynamics of free quantum scalar fluctuations over the orthogonal surfaces to both strings. The factorisation properties of the heat kernel allow to obtain straight forward the heat kernel coefficients for the whole system including the space dimension parallel to the strings.

\subsection{Geometrical formulas for 2 infinitely thin cosmic strings}
The spatial metric tensor for two cosmic strings in a space of dimension $D$ is
\begin{equation}
 g=e^{-4 V}\times g^{(D)}_{{\rm flat}}
\end{equation}with
\begin{equation}
V=2\sum_{i=1}^2\lambda_i\log (r_i)=V_2 +V_2,\quad r_i=\vert {\bf r}-{\bf R}_i\vert
\end{equation}and
\begin{equation}
{\bf R}_i=((-1)^{i+1}d,0,...,0) \quad (i=1,2).
\end{equation}
In this case the general geometrical formulas simplify,
\begin{equation}
\Gamma_{ij}^k=0,\quad {\rm Riem}=0\Rightarrow \nabla_k=\partial_k,
\epsilon=1,\quad F=-2 V.
\end{equation}
Tensorial objects arising from the derivatives of $V$ are
\bea \pa_j V&=& 2(x_j+\delta_{j1}d)\frac{\lambda_1}{r_1^2}+2(x_j-\delta_{j1}d)\frac{\lambda_2}{r_2^2},
\nn \\
\partial_j\partial_kV &=& -\frac{4\lambda_1}{r_1^4}(x_k+\delta_{1k}d)(x_j+\delta_{1j}d)-\frac{4\lambda_2}{r_2^4}(x_k-\delta_{1k}d)(x_j-\delta_{1j}d)\nonumber\\
&+&2\delta_{jk}\left( \frac{\lambda_1}{r_1^2}+\frac{\lambda_2}{r_2^2}\right),
\nn\\
\partial_jV\partial_kV&=&\frac{4\lambda_1^2}{r_1^4}(x_k+\delta_{1k}d)(x_j+\delta_{1j}d)+\frac{4\lambda_2^2}{r_2^4}(x_k-\delta_{1k}d)(x_j-\delta_{1j}d)\nonumber\\
&+&\frac{4\lambda_1\lambda_2}{r_1^2r_2^2}\left((x_k+\delta_{1k}d)(x_j-\delta_{1j}d)+(x_k-\delta_{1k}d)(x_j+\delta_{1j}d)\right)
.\eea
Using these tensors we can compute for arbitrary spatial dimension $D$ the Laplacian contraction  and the gradient contraction,
\bea
\partial_k\partial^kV &=&2(D-2)\left( \frac{\lambda_1}{r_1^2}+\frac{\lambda_2}{r_2^2}\right)   \quad (D\neq 2),        \nn\\
\partial_jV\partial^jV &=& \frac{4\lambda_1^2}{r_1^2}+\frac{4\lambda_2^2}{r_2^2}+\frac{8\lambda_1\lambda_2}{r_1^2 r_2^2}({\bf r}_1\cdot{\bf r}_2).
\eea
The Laplacian of $V$ for $D=2$ must be computed carefully because $\log (r)$ is an harmonic function in dimension 2,
\begin{equation}
\lim_{D\to 2}\partial_k\partial^k\log(\vert{\bf r}-{\bf R}\vert)=2\pi\delta^{(2)}({\bf r}-{\bf R}).
\end{equation}
Hence using the harmonic equation   we obtain for $D=2$
\begin{equation}
\lim_{D\to 2}\partial_k\partial^kV=4\pi\left(\lambda_1\delta^{(2)}({\bf r}-{\bf R}_1)+\lambda_2\delta^{(2)}({\bf r}-{\bf R}_2)\right).
\end{equation}

With these we can write down
\bea
e^{-8V}\widetilde\Delta\widetilde R&=&-16(D-1)(3D-10)(\partial_nV\partial^nV)(\partial_k\partial^kV)+16(D-1)(\partial_k\partial^kV)^2 \nonumber\\
&&+16(D-1)(D-2)\left(2(D-4)(\partial_nV\partial^nV)^2+(D-6)\partial_k\partial_jV\partial^k\partial^jV\right)\nonumber\\
&&+4(D-1)\partial_n\partial^n\partial_k\partial^kV-8(D-1)(D-2)\partial_j\partial_kV\partial^j\partial^kV\nonumber\\
&&-(D-1)(D-2)\partial^kV\partial_k(\partial_n\partial^nV)-8(D-1)(D-6)\partial^kV\partial_k(\partial_n\partial^nV)\label{lap-curv}
,\nn\\
e^{-8V}\widetilde{R}^2 &=&16(D-1)^2(\partial_n\partial^nV)^2-32(D-1)^2(D-2)(\partial_nV\partial^nV)(\partial_k\partial^kV)\nonumber\\
&&+(D-1)^2(D-2)^2(\partial_n\partial^nV)^2\label{curv},
\nn\\
e^{-8V}\widetilde R^{ij}\widetilde R_{ij}&=&4(D-2)^2\partial_i\partial_jV \partial^i\partial^jV+4(3D-2)(\partial_n\partial^nV)^2 \nonumber\\
&&+16(D-2)(3-2D)(\partial_kV\partial^kV)(\partial_n\partial^nV)+16(D-1)(D-2)^2(\partial_kV\partial^kV)^2\nonumber\\
&&+16(D-2)^2\partial_jV\partial_kV\partial^j\partial^kV\label{ricci-sq},
\nn\\
e^{-8V}\widetilde R^{ijkl}\widetilde R_{ijkl}&=&16(D-2)\partial_j\partial_kV\partial^j\partial^kV+64(D-2)\partial^jV\partial^kV\partial_j\partial_kV \nonumber\\
&&-64(D-2)(\partial_nV\partial^nV)(\partial_k\partial^kV)+16(\partial_n\partial^nV)^2\nonumber\\
&&+32(D-1)(D-2)(\partial_nV\partial^nV)^2.
\eea\label{riem-sq}

\section{The heat kernel coefficient $a_4$ is zero for two cosmic strings }
In order to demonstrate that there are no divergences in the quantum vacuum interaction energy between two cosmic strings we must demonstrate that the heat kernel coefficient $a_4$ is identically zero. Taking into account the geometrical formulas given in the last section and the general expression for $a_4$, the integrand will have a global factor $\sqrt{g}e^{8V}$. Using the expressions for $V$ and $g$ for two cosmic strings is easy to notice that
\begin{equation}
\sqrt{g}e^{8V}=r_1^{8\lambda_1}r_2^{8\lambda_2}.
\end{equation}
The presence of this global factor does not allow to split the heat kernel coefficient $a_4$ into 3 terms,
\begin{equation}
a_4=a_4^{(1\,string)}+a_4^{(1\,string)}+a_4^{(int)}.
\end{equation}
Hence we must go over the whole computation for the heat kernel coefficient $a_4$.

To start with the explicit calculation of $a_4$, the first step is to drop those terms appearing in the integrand of the geometric expression for $a_4$ that are proportional to $D-2$. These terms will give zero contribution when taking the physical limit $D\rightarrow 2$. Collecting the remaining terms from each geometrical combination appearing in \Ref{gena4} and using equations \Ref{lap-curv} we obtain
\bea
e^{-8V}\widetilde\Delta\widetilde R&=&-16(D-1)(3D-10)(\partial_nV\partial^nV)(\partial_k\partial^kV)+16(D-1)(\partial_k\partial^kV)^2 \nonumber\\
&&+4(D-1)\partial_n\partial^n\partial_k\partial^kV -8(D-1)(D-6)\partial^kV\partial_k(\partial_n\partial^nV)+ \mathcal{O}(D-2),
\nn\\
e^{-8V}\widetilde{R}^2 &=&16(D-1)^2(\partial_n\partial^nV)^2+ \mathcal{O}(D-2),
\nn\\
e^{-8V}\widetilde R^{ij}\widetilde R_{ij}&=&4(3D-2)(\partial_n\partial^nV)^2 + \mathcal{O}(D-2),
\nn\\
e^{-8V}\widetilde R^{ijkl}\widetilde R_{ijkl}&=&16(\partial_n\partial^nV)^2+ \mathcal{O}(D-2).
\eea
Note that now   the $a_4$ coefficient will have the form
\begin{equation}
 (4\pi)^{D/2}a_4=\int_M D^{D}x\sqrt{g}e^{8V}p_1\left((\partial_kV)^2\partial^2 V,(\partial^2V)^2,\partial^2\partial^2V,\partial^kV\partial_k(\partial^2V))\right)+\mathcal{O}(D-2),
\end{equation}
where $p_1(x,y,z,t)$ is a polynomial of order 1 in four variables whose independent term is zero. This polynomial is obtained by adding the expressions given above after dropping terms proportional to $D-2$ with the coefficients given in \Ref{gena4}. For our purpose   the values of the coefficients appearing in $p_1$ are not important. Nevertheless note that the whole integrand is proportional to
\begin{equation}
e^{8V}\sqrt{g}=r_1^{8\lambda_1}r_2^{8\lambda_2}.
\end{equation}
As was said before this term ensures that all the terms in $a_4$ are influenced by the interaction of both strings. But if we also take into account that
\begin{equation}
\lim_{D\rightarrow2}\partial^2V \sim \lambda_1\delta^{(2)}({\bf r}_1)+\lambda_2\delta^{(2)}({\bf r}_2)
\end{equation}
and the formulas for $\partial_kV$ then
\begin{equation}
\lim_{D\rightarrow2}\int_M d^D x\sqrt{g}  e^{8V}(\partial_kV)^2\partial^2 V=0.
\end{equation}
Integrations concerning the terms $(\partial^2V)^2$, $\partial^2\partial^2V$,  and $\partial^kV\partial_k(\partial^2V)$ can be reduced to the one above by integrating by parts and give zero contribution to the heat kernel coefficient $a_4$ when taking the limit $D\rightarrow 2$. Therefore
\begin{equation}
\lim_{D\rightarrow 2}a_4^{2strings}(\lambda_1,\lambda_2,D)=0,
\end{equation}
which ensures the finiteness of the quantum vacuum interaction energy between two cosmic strings.

\section{The full calculation of the quantum vacuum interaction between two cosmic strings}
The general formula for the vacuum energy is given by eq. (3.112) in reference \cite{BKMM} and after Fourier transform in the translational invariant directions it can be written as
\begin{equation}
    E_0=-\frac{i}{2}\int\frac{dk_0dk_3}{(2\pi)^2} \
    {\rm Tr}\log\mathcal{G}_{\gamma},  \label{5.1}
\end{equation}
where $\mathcal{G}_{\gamma}$ is the exact Green function \Ref{green-1cs} for the quantum fluctuations in the corresponding classical background. In \Ref{5.1} the trace is over the coordinates in the section perpendicular to the string. Following now section 4 in reference \cite{bord12-45-374012} we can write
\begin{equation}
\mathcal{G}_{\gamma}=\frac{\mathcal{G}^{(0)}_{\gamma}}{{\boldsymbol 1}-\mathcal{V}\cdot \mathcal{G}^{(0)}_{\gamma}},
\end{equation}
being $\mathcal{V}$ the potential operator characterising the classical background. Doing in addition the Wick rotation $k_0\to ik_4$ and using rotational invariance in the $(k_4,k_3)$-plane  with now $\gamma=\sqrt{k_4^2+k_3^2}$, we can write
\begin{equation}
E_0=E_{empty}-\frac{1}{2}\int_m^\infty\frac{d\gamma}{2\pi}\,\gamma \, {\rm Tr}\log({\boldsymbol 1}-\mathcal{V}\cdot \mathcal{G}^{(0)}_{\gamma}),
\end{equation}
where  $E_{empty}$ is the infinite valued empty space vacuum energy
\begin{equation}
E_{empty}= \frac{1}{2}\int_m^\infty\frac{d\gamma}{2\pi}\,\gamma\,{\rm Tr}\log\mathcal{G}^{(0)}_{\gamma}.
\end{equation}
What has physical meaning is the quantity $E_{V}=E_0-E_{empty}$ that is now given by
\begin{equation}
E_V=-\frac{1}{2}\int_0^\infty\frac{d\gamma}{2\pi}\,\gamma\,{\rm Tr}\log({\boldsymbol 1}-\mathcal{V}\cdot \mathcal{G}^{(0)}_{\gamma}).
\end{equation}
For the two parallel cosmic strings system the potential is given by
\begin{equation}
\mathcal{V}=e^{-4(V_1+V_2)}(\partial^2_{||}+m^2),
\end{equation}
being $\partial^2_{||}=\partial_3^2$. These equation will allow us to compute
\begin{enumerate}
\item the quantum vacuum interaction energy when both strings are accounted for perturbatively,
\item the quantum vacuum interaction energy when one of the strings is accounted for non-perturbatively and the other perturbatively.
\end{enumerate}
\subsection{The first order contribution to the vacuum interaction energy}
The approach we will take to compute the quantum vacuum interaction energy between two cosmic strings will be simpler than the one taken in reference \cite{Bordag:1990if}: we will account for both mass densities only up to first order. The difference is that the calculation will be done in momentum space using the 'Trace log' formula \Ref{5.1} for the vacuum energy in place of the energy-momentum tensor. Taking into account formula (3.112) in reference \cite{BKMM} we can write the quantum vacuum interaction energy between two cosmic strings up to first order in $\lambda_1$ and $\lambda_2$ as a functional trace
\begin{equation}
E_0^{(1)}(b,\lambda_1,\lambda_2)=\frac{16}{2} {\rm Tr}\left(\mathcal{V}_1\cdot \mathcal{G}_0\cdot \mathcal{V}_2\cdot \mathcal{G}_0\right),    \label{5.7a}
\end{equation}
where $G_0$ is the empty flat space free propagator, $\mathcal{V}_i=V_i(-\partial_{||}^2)$ with $V_i=-2\lambda_i\log(r_i)$, and $\partial_{||}$ being the derivative with respect the spatial coordinate parallel to the strings. In order to simplify the calculation we will compute the functional trace over the momentum space. That means that we must write the empty space free propagator as
\begin{equation}
G_0(z)=\int\frac{d^4p}{(2\pi)^4}\frac{e^{ip\cdot z}}{p_\mu p^\mu}.
\end{equation}
On the other hand side we must write the potential operator $V_i$ in the momentum space. Since the potential contains a derivative, the first step is to write down the action of $V_i$ as an operator over the empty space-time Green function,
\begin{equation}
\mathcal{V}_i\cdot G_0(x-y)=V_i\int\frac{d^4p}{(2\pi)^4}\frac{p_{||}^2e^{ip\cdot (x-y)}}{p_\mu p^\mu}.
\end{equation}
Note that $p^2=p_{||}^2+p_{\perp}^2$. Now we can write down the Fourier transform of the factors $V_i$ in order to eliminate all spatial integrations and end up with a multiple integration over momenta. It is not difficult to check that
\begin{equation}
\tilde{V}_i(k)=(2\pi)^2\delta(k_0)\delta(k_{||})\frac{4\pi}{k_{\perp}^2} \lambda_i e^{i{\bf k}_{\perp}\cdot{\bf s}_i}.
\end{equation}
Hence now the kernel of the operator product $\mathcal{V}\mathcal{G}_0$ can be written as
\begin{equation}
\left(\mathcal{V}_i\mathcal{G}_0\right)(x,y)=
\int\frac{d^4k}{(2\pi)^4}\frac{d^4p}{(2\pi)^4}\tilde{V}_i(k)\,e^{ikx}\,\frac{p_{||}^2}{p^2}\,e^{ip\cdot (x-y)}.
\end{equation}
Therefore the functional trace ${\rm Tr}\left(\mathcal{V}_1\cdot \mathcal{G}_0\cdot \mathcal{V}_2\cdot \mathcal{G}_0\right)$ can now be written in terms of integrations over the momenta and we get for \Ref{5.7a}
\begin{equation}
E_0=\frac{16}{2}
\int\frac{d^4k}{(2\pi)^4}\frac{d^4p}{(2\pi)^4}V(k)\frac{p_{||}^2}{p^2}V(-k)\frac{(k+p)_{||}^2}{(k+p)^2}
.\end{equation}
Using the explicit expressions for the Fourier transform of the potentials we can integrate out $k_{||}$ and write down the expression for the energy density,
\begin{equation}
{E_0}=128\lambda_1\lambda_2\int \frac{d^2k_{\perp}}{(2\pi)^2}\frac{e^{i{\bf k}_{\perp}\cdot{\bf b}}}{k_\perp^2}\Sigma({\bf k}_\perp),
\end{equation}where we droped the volume factor $VT$ and introduced the notation
\begin{equation}
\Sigma({\bf k}_\perp)=\int\frac{d^4p}{(2\pi)^4}\frac{(p_{||}^2)(k+p)_{||}^2}{(k+p)^2}\Big|_{{\bf k}_{||}=0}.
\end{equation}
The calculation of $\Sigma({\bf k}_\perp)$ requires the introduction of dimensional regularisation,
\begin{equation}
\Sigma({\bf k}_\perp,\epsilon)=\left.\int\frac{d^{4-2\epsilon}p}{(2\pi)^{4-2\epsilon}}\frac{(p_{||}^2)(k+p)_{||}^2}{(k+p)^2}\right]_{{\bf k}_{||}=0}.
\end{equation}
The regularised integration $\Sigma({\bf k}_\perp,\epsilon)$ can be exactly computed and is given by
\begin{equation}
\Sigma({\bf k}_\perp,\epsilon)=
\frac{\Gamma(\epsilon)\Gamma(3-\epsilon)^2}{(4\pi)^{4-2\epsilon}
\Gamma(2(3-\epsilon))}k_\perp^{4-2\epsilon}
\equiv g_\epsilon k_\perp^{4-2\epsilon}.
\end{equation}
In this way we obtain a regularised expression for the vacuum interaction energy density,
\begin{equation}
{E_0}=128 \pi^2 \lambda_1\lambda_2 \,g_\epsilon\int \frac{d^2k_{\perp}}{(2\pi)^2}\,e^{i{\bf k}_{\perp}\cdot{\bf b}}k_\perp^{-2\epsilon}.
\end{equation}
The regularised integral in the last expression can be exactly computed as a function of $\epsilon$,
\begin{equation}
\int \frac{d^2k_{\perp}}{(2\pi)^2}e^{i{\bf k}_{\perp}\cdot{\bf b}}k_\perp^{-2\epsilon}
=\frac{-\epsilon}{4\pi}\left(\frac{2}{b}\right)^{2(1+\epsilon)}.
\end{equation}
In order to take the limit $\epsilon\rightarrow 0$ it is important to note that
\begin{equation}
\epsilon g_\epsilon=\frac{1}{30\cdot(4\pi)^2}+\mathcal{O}(\epsilon),
\end{equation}
thus the removal of the regularization results in an ultraviolet finite quantity.
In this way, the perturbative calculation for the quantum vacuum interaction energy density between two parallel cosmic strings results in
\begin{equation}
 {E_0} =-\frac{4\lambda_1\lambda_2}{15\pi}\left.\frac{1}{b^2}\right..
\end{equation}
Even when this perturbative calculation is correct it might not account for nonperturbative  geometrical  contributions. Therefore one could expect contributions of geometrical nature that are of order $\lambda_1\lambda_2$ that can not be computed in this approach.
Hence it is worth to compute the quantum vacuum interaction energy in a way that these geometrical terms can be studied. This is a motivation for the next subsection where, like in \cite{Bordag:1990if}, one of the cosmic strings is accounted for in a non perturbative way, and whereas the second string is a potential defined over the space-time defined by the first string. This means that the second string would be accounted perturbatively.

\subsection{Quantum vacuum interaction energy when one string is accounted non-perturbatively}
In this section we calculate the vacuum  energy taking the first string exactly and the second perturbatively. The vacuum energy is given by
\be\label{5.1a}
    E_0=-2         \int\frac{d^{2-2\ep}k}{(2\pi)^{2-2\ep}}\,
    {\rm Tr}V_2\rho^2\gamma^2G_{\gamma}(\xes,\xes'),
\ee
where we introduced dimensional regularization by adding dimensions in the directions parallel to the strings which have now dimension $d=2(1-\ep)$. In \Ref{5.1a},
\be\label{5.2}V_2=2\lambda_2\ln R=2\lambda_2\ln\sqrt{r^2+b^2-2b r \cos(\varphi)}
\ee
is the potential of the second string which we split into two parts according to
\be\label{5.3}V_2=\ln b+\ln \tilde{R}
\ee
with
\be\label{5.4}\tilde{R}=\sqrt{\left(\frac{r}{b}\right)^2+1-2\frac{r}{b}\cos(\varphi)}.
\ee
In eq. \Ref{5.1a}, the trace is
\be\label{5.5}
        {\rm Tr}=\int\limits_0^\infty\frac{d\rho}{\rho}
        \int\limits_0^{2\pi}d\varphi.
\ee
 In eq. \Ref{5.1a}, the Green function is given by eq. \Ref{green-1cs}, and we get
\be\label{5.6}
    E_0=-4\lambda_2\int\limits_0^\infty\frac{d\rho}{\rho}\int\limits_0^{2\pi}d\psi
    \rho^2(\ln b+\ln\tilde{R})\int\frac{d^{2-2\ep}k}{(2\pi)^{2-2\ep}}
    \gamma^2 \int\limits_0^\infty %
    d\mu \ K_{i\mu}(\gamma\rho)^2
    G_\mu(0)
\ee
with
\be\label{5.7}
   G_\mu(0)=  \frac{1}{2\mu}\coth(\alpha\pi\mu)
\ee
following from \Ref{green-angular} with $\Delta\psi=0$. Now we consider the first contribution to $E_0$ according to the splitting in eq. \Ref{5.3} or, equivalently, the contribution from the first term in the parentheses in \Ref{5.6}. Since $\ln b$ does not depend on $\rho$, we can make the substitution $\rho\to\rho/\gamma$ after which the momentum dependence drops out and the pure integration over $k$ from the trace, eq. \Ref{5.5}, remains. This integration appears to be not regularized by the dimensional regularization, thus requiring some additional regularization, a momentum cut-off $\Lambda$ for instance. Then the integral becomes proportional to $\Lambda^2$, carrying the dimension of the expression. As a result the separation dependence is in the logarithm only. Since it does not decrease for $b\to\infty$ we have to drop this contribution completely. In this way we are left with the second term in the parenthesis in eq. \Ref{5.6}.
Next we focus on the massless case. This allows for the substitution $k\to k/\rho$ and the expression for the energy factorizes into a product of two,
\be\label{5.8}  E_0=-4\lambda_2 C_1 C_2
\ee
where
\be\label{5.9}  C_1=\int\limits_0^\infty\frac{d\rho}{\rho}\int\limits_0^{2\pi}d\psi \
                \frac{1}{\rho^2}\ln \tilde{R}
\ee
has the spatial integrations perpendicular to the strings and
\be\label{5.10} C_2=\int\frac{d^{2-2\ep}k}{(2\pi)^{2-2\ep}}
    k^2 \int\limits_0^\infty %
    d\mu \ K_{i\mu}(k)^2
    g(\mu)
\ee
has the remaining integrations. First, we consider $C_2$. Here the momentum integration can be carried out. With spherical coordinates for $k$ we get
\be\label{5.11}
    C_2=\frac{\Omega_{2-2\ep}}{\pi^2(2\pi)^{2-2\ep}} \int\limits_0^\infty
    d\mu\ g(\mu)\int\limits_0^\infty dk\ k^{3-2\ep}K_{i\mu}(k)^2,
\ee
where $\Omega_n=2\pi^{n/2}/\Gamma(n/2)$ is the surface of the $n$-dimensional sphere. The integration over $k$ can be done using eq. (6.5764) in \cite{grad94},
\be\label{5.12}
    \int\limits_0^\infty dk\ k^{3-2\ep} K_{i\mu}(k)^2=\frac{\sqrt{\pi}\,\Gamma(2-\ep)}{4\Gamma(\frac52 -\ep)}\ h(\ep,\mu)
\ee
with
\be\label{5.13} h(\ep,\mu)=\Gamma(2-\ep+i\mu)\Gamma(2-\ep-i\mu).
\ee
This function has the following properties,
\bea\label{5.14} h(0,\mu)&=&\frac{\pi\mu(1+\mu^2)}{\sinh(\pi\mu)},  \nn \\
        h(\ep,\mu)&=&2\pi\mu^{3-2\ep}e^{-\pi\mu}\left(1+\frac{a_1(\ep)}{\mu^2}+\frac{a_2(\ep)}{\mu^4}
        +O\left(\frac{1}{\mu^6}\right)\right),
\eea
where the first is the function for $\ep=0$ and the second is the asymptotic expansion for $\mu\to\infty$ which is obtained using Stirling's formula. Special values of its coefficients are
\be\label{5.15} a_1(0)=1,\quad a_2(\ep)=-\frac{11}{60}\ep+O(\ep^2).
\ee
The $\mu$-integration in \Ref{5.11} does not converge on the upper boundary if putting $\ep=0$ under the sign of the integration, as can be seen from eqs. \Ref{5.7} and \Ref{5.14}. However, for $\ep>2$ the integration is convergent. Therefor we need to calculate the analytic continuation of $C_2$ from $\ep>2$ to $\ep=0$. This can be done in the following way. First, we rewrite $C_2$ in the form
\be\label{5.16} C_2=d_\ep \int\limits_0^\infty  d\mu \ h(\ep,\mu)\,g(\mu),
\ee
where
\be\label{5.16a}d_\ep=\frac{\Omega_{2-2\ep}}{\pi^2(2\pi)^{2-2\ep}}
                \frac{\sqrt{\pi}\,\Gamma(2-\ep)}{4\Gamma(\frac52 -\ep)}
\ee
collects the constants in front with the special case
\be\label{5.16b}d_0=\frac{1}{6\pi^3}.
\ee
Next we rewrite $g(\mu)$, eq. \Ref{5.7}, in the form $g(\mu)=\sinh(\pi\mu)\left[\coth(\alpha\pi\mu)-\coth(\pi\mu)+\coth(\pi\mu)\right]$
and split accordingly,
\be\label{5.17} C_2=I_0+I_1,
\ee
with
\bea\label{5.18}    I_0&=&d_\ep \int\limits_0^\infty  d\mu \ h(\ep,\mu)
    \sinh(\pi\mu)\left[\coth(\alpha\pi\mu)-\coth(\pi\mu)\right],  \nn\\
         I_1&=&d_\ep \int\limits_0^\infty  d\mu \ h(\ep,\mu)
    \sinh(\pi\mu)\coth(\pi\mu)  .
\eea
In $I_0$, due to the compensation between the two hyperbolic cotangents, we can put $\ep=0$ directly and get a convergent integral, which can be rewritten in the form
\be\label{5.19} I_0=\frac{1}{3\pi^2} \int\limits_0^\infty  d\mu \ \mu(1+\mu^2)
        \left(\frac{1}{e^{2\alpha\pi\mu}-1}-\frac{1}{e^{2\pi\mu}-1}\right),
\ee
allowing for an easy integration  resulting in
\be\label{5.20} I_0=\frac{(1-\alpha^2)(1+11\alpha^2)}{720\pi^2\alpha^4}.
\ee
The expression for $I_1$ can be rewritten in the form
\be\label{5.21} I_1=\frac{d_\ep}{2} \int\limits_0^\infty  d\mu \
                h(\ep,\mu)\left(e^{\pi\mu}-e^{-\pi\mu}\right).
\ee
Here the second part converges and the divergence is in the first part. We introduce the notations
\be\label{5.22} I_1=P+Q
\ee
with
\bea\label{5.23} P&=&\frac{d_\ep}{2} \int\limits_0^\infty  d\mu \ h(\ep,\mu) e^{\pi\mu},\nn\\
                Q&=&\frac{d_\ep}{2} \int\limits_0^\infty  d\mu \ h(\ep,\mu) e^{-\pi\mu}.
\eea
In $Q$ we can put $\ep=0$ directly and get
\be\label{5.24} Q=\frac{1}{12\pi^2}\int\limits_0^\infty  d\mu \
                \mu(1+\mu^2)\frac{e^{-\pi\mu}}{\sinh(\pi\mu)}=\frac{11}{1440\pi^2}.
\ee
Now we turn to $P$. In order to construct the continuation in $\ep$,  we use the asymptotic expansion \Ref{5.14} and define
\be\label{5.25} h_{\rm as}(\ep,\mu)=
    2\pi\mu^{3-2\ep}e^{-\pi\mu}\left(1+\frac{a_1(\ep)}{\mu^2}+\frac{a_2(\ep)}{\mu^4}\right).
\ee
We mention the property
\be\label{5.25a} h_{\rm as}(0,\mu)=2\pi\,\mu(1+\mu^2)e^{-\pi\mu}.
\ee
This allows to rewrite $P$, eq. \Ref{5.23}, in the form
\be\label{5.26}
    P=\frac{d_\ep}{2} \left[ \int\limits_0^1  d\mu \ h(\ep,\mu)e^{\pi\mu}
                +\int\limits_1^\infty d\mu\ \left(h(\ep,\mu)-h_{\rm as}(\ep,\mu)\right)e^{\pi\mu}
                +\int\limits_1^\infty d\mu\  h_{\rm as}(\ep,\mu) e^{\pi\mu}       \right].
\ee
Now, in the first two integrals, due to their convergence, we can put $\ep=0$ directly and get
\bea\label{5.27}
   && \frac{d_\ep}{2} \left[ \int\limits_0^1  d\mu \ h(\ep,\mu)
                +\int\limits_1^\infty d\mu\ \left(h(\ep,\mu)-h_{\rm as}(\ep,\mu)\right)\right]\nn\\
   &&=\frac{1}{12\pi^3} \left[ \int\limits_0^1  d\mu \ h_{\rm as}(0,\mu)
                +\int\limits_0^\infty d\mu\ \left(h(0,\mu)-h_{\rm as}(0,\mu)\right)\right]\nn\\
   &&=\frac{191}{1440\pi^2},
\eea
where we used \Ref{5.14} and \Ref{5.25a}. The last integration was similar to eq. \Ref{5.24}.
Further, the last integral in \Ref{5.26} can be written as
\be\label{5.28} P=\frac{191}{1440\pi^2}+\frac{d_\ep}{2}\int\limits_1^\infty d\mu\ 2\pi\left(\mu^{3-2\ep}+a_1(\ep)\mu^{1-2\ep}+a_2(\ep)\mu^{-1-2\ep}\right),
\ee
where \Ref{5.25} was used. This integration is now also easy,
\be\label{5.29} P=\frac{191}{1440\pi^2}+ d_\ep \pi
     \left(\frac{-1}{4-2\ep}-\frac{a_1(\ep)}{2-2\ep}-\frac{a_2(\ep)}{-2\ep}\right).
\ee
After the integration  carried out we can make the continuation to $\ep=0$. Using \Ref{5.15} we get
\be\label{5.30} P=\frac{191}{1440\pi^2}-\frac{101}{720\pi^2}=-\frac{11}{1440\pi^2}.
\ee
The disappearance of a singularity for $\ep\to0$ comes in since $a_2(\ep) $ is proportional to $\ep$, see eq. \Ref{5.15}.

Now we return to \Ref{5.22} and insert \Ref{5.24} and \Ref{5.30}, which compensate each other,
\be\label{5.31} I_1=0.
\ee
Further, returning to eq. \Ref{5.17}, we get with \Ref{5.20} for the continuation to $\ep=0$,
\be\label{5.32} C_2=\frac{(1-\alpha^2)(1+11\alpha^2)}{720\pi^2\alpha^4},
\ee
which concludes the calculation of $C_2$.

In order the calculate $C_1$, eq. \Ref{5.9}, we use \Ref{5.4} and for the angular integration we use
\be\label{5.33} \int\limits_0^{2\pi} d\varphi\ \ln\sqrt{x^2+1-2x\cos(\varphi)}=
    2\pi\ln(x)\Theta(x-1).
\ee
Also we need the relation between conformal and flat radii, $\rho=\frac{r^\alpha}{\alpha}$. We get
\bea\label{5.34} C_1&=&\int\limits_0^\infty\frac{d\rho}{\rho}\int\limits_0^{2\pi} d\varphi\
                \frac{1}{\rho^2}\ln\sqrt{\left(\frac{r}{b}\right)^2+1-2\frac{r}{b}\cos(\psi)}
\nn\\           &=&2\pi\int\limits_b^\infty \frac{d r}{r}\frac{\alpha^2}{r^{2\alpha}}\ln\left(\frac{r}{b}\right)
\eea
resulting in
\be\label{5.35} C_1=\frac{\pi}{2\,b^{2\alpha}}.
\ee
Finally, returning to eq. \Ref{5.8}, we get for the vacuum energy
\be\label{5.36} E_0=-\lambda_2\frac{(1-\alpha^2)(1+11\alpha^2)}{720\pi \alpha^4 \,b^{2\alpha}},
\ee
which is the final answer for the vacuum energy if one string is taken exactly and the second perturbatively.

As a special case we get with $\alpha=1-4\lambda$ for $\lambda\to0$ the case if both strings are taken perturbatively,
\be\label{5.37} E_0=-\frac{\lambda\lambda_2}{b^2}\,\frac{4}{15\pi}.
\ee
This result coincides with eq. (35) in \cite{galt95-58-3903} and eq. (34) in \cite{alie97-55-3903} (after dividing by 2 for the polarizations of the electromagnetic field).

\section{Alternative model for non interacting quantum cosmic strings:  delta function potentials in two dimensions}
The main papers  and notes used to develop the TGTG approach to the vacuum interaction between two delta functions in two diemsnions are
\begin{itemize}
\item R. Jackiw: Delta function potentials in two and three dimensional quantum mechanics \cite{jackiw95},
\item P. Gosdzinski and R. Tarrach: Learning quantum field theory from elementary quantum mechanics \cite{godz91-59-70}. This paper deals also with the delta potential in 2D,
\item I. R. Lapidus: Quantum mechanical scattering in two dimensions \cite{lapi82-50-45}. This paper is about 2D scattering theory and also deals with the delta potential in 2D,
\end{itemize}
One of the main classical aspects that characterise cosmic strings is the fact that at the classical level they do not interact. As has been seen  from quantum corrections to the interaction between cosmic strings it emerges a non zero interaction. Therefore since the classical interaction is identically zero the quantum interaction is present like in the original Casimir effect.
Here we consider as another example strings carrying two dimensional delta functions. There is no  classical interaction and the quantum interactions, say of a scalar field, in that background was never considered. The configuration of two 'Delta' strings   can be viewed parallel to that of two planes carrying delta function potentials (one dimensional delta functions in that case) showing a non zero Casimir force \cite{Bordag:1992cm,cast13-87-105020}. Below we will  demonstrate that the quantum interaction between such strings is zero when their thickness is zero.

\subsection{General formulas for TGTG in in two dimensions}
The free Green function in  the flat 2-dimensional space is given by
\begin{equation}
G^{(0)}_\omega ({\bf r}-{\bf r} ')=\frac{i}{4} H^{(1)}_0(\omega |\bf{r}-\bf{r} '|),
\end{equation}
where $H^{(1)}_m(z)$ are the 1$^{\rm st}$ kind Hankel function of $m$-th order. The Born series for the $T$-operator is given by (see reference \cite{BKMM})
\begin{eqnarray}
{\bf T}_\omega & = & {\bf V}+{\bf V}\cdot{\bf  G}_\omega^{(0)}\cdot {\bf V}+{\bf V}\cdot {\bf G}_\omega^{(0)}\cdot {\bf V}\cdot {\bf G}_\omega^{(0)}\cdot {\bf V}\nonumber \\
&+& {\bf V}\cdot {\bf G}_\omega^{(0)}\cdot {\bf V}\cdot {\bf G}_\omega^{(0)}\cdot {\bf V}\cdot {\bf G}_\omega^{(0)}\cdot {\bf V}+...=(1+{\bf V}\cdot {\bf G}_\omega^{(0)})^{-1}\cdot {\bf V}.
\end{eqnarray}
Hence for a    delta potential we can write for any dimensionality the following formal expression for the kernel of the $T$-operator,
\begin{equation}
T_\omega({\bf x},{\bf x}';\,\alpha)=\frac{\alpha\delta^{(n)}({\bf x})\delta^{(n)}({\bf x}')}{1-\alpha G_\omega^{(0)}({\bf x}-{\bf x}')},
\end{equation}
where $\alpha$ is the weight of the  delta potential and $G_\omega^{(0)}$ is the free Green function in $n$-dimensional flat space.

\subsection{TGTG formula for the interaction between two delta functions in two dimensions}
The potential for a single two dimensional delta function is: $V({\bf r})=\alpha\delta^{(2)}({\bf r})$. This potential is not well defined in the 2-dimensional case (see references \cite{jackiw95,lapi82-50-45,godz91-59-70}) and requires the use of an energy cutoff. In references \cite{jackiw95,godz91-59-70} it is demonstrated that this potential is asymptotically free due to the fact that the coupling constant is dimensionless. To compute the quantum vacuum interaction between two Dirac delta strings we will introduce a regularisation parameter with physical meaning: the thickness of the Dirac Delta strings\footnote{For simplicity we assume that both Dirac delta strings have the same thickness $r_0$.}. This will be done by introducing a minimal radius ${\bf r}_0$ in the free Green function. By doing this we assume that the thickness of the Dirac delta string is nothing but $r_0$

From the Born series and the Lipmann-Schwinger equation we obtain the $T$-operator for a single $\delta$ at the origin:
\begin{equation}
T_\omega({\bf r},\,{\bf r}';\,\alpha)=\frac{\alpha\delta^{(2)}({\bf r})\delta^{(2)}({\bf r}')}{1+\alpha G^{(0)}_\omega({\bf r}-{\bf r}')}
\end{equation}
Since $H^{(1)}_0(z)$ has a logarithmic divergence when $z\to 0$ we must introduce at this point a regularisation ${\bf r}_0$ that has the physical meaning of the thickness of the Dirac delta string,
\begin{equation}
T_\omega({\bf r},\,{\bf r}';\,\alpha,\,{\bf r}_0)=\frac{\alpha\delta^{(2)}({\bf r})\delta^{(2)}({\bf r}')}{1+\alpha G^{(0)}_\omega({\bf r}_0+{\bf r}-{\bf r}')}.
\end{equation}
Our system contains two  2D delta functions, one at the origin (weight $\alpha$), and the other at ${\bf a}$ (weight $\beta$). The $T$-operator for the displaced delta function is obtained by just changing ${\bf r}$ and ${\bf r}'$ for ${\bf r}-{\bf a}$ and ${\bf r}'-{\bf a}$. Hence the kernel of the operator ${\bf M}_\omega$ (see reference \cite{BKMM}), given by
\begin{eqnarray*}
M_\omega({\bf r},{\bf r}')&=&\int d{\bf z}_1d{\bf z}_2d{\bf z}_3 G^{(0)}_\omega({\bf r}-{\bf z}_1)T_\omega({\bf z}_1,\,{\bf z}_2;\,\alpha,\,{\bf r}_0)\\
& &\times G^{(0)}_\omega({\bf z}_2-{\bf z}_3)T_\omega({\bf z}_3-{\bf a},\,{\bf r}'-{\bf a};\,\beta,\,{\bf r}_0),
\end{eqnarray*}
can be easily computed by using in the regularised expression for the corresponding $T$-operators,
\begin{equation}
M_\omega({\bf x},{\bf x}';\,\alpha,\beta;\,{\bf r}_0)=\alpha\beta\frac{G^{(0)}_\omega({\bf x})\delta^{(2)}({\bf x}'-{\bf a}) G^{(0)}_\omega(-{\bf a})}{(1+\alpha G^{(0)}_\omega({\bf r}_0))(1+\beta G^{(0)}_\omega({\bf r}_0+{\bf a}-{\bf x}'))}.
\end{equation}
The lowest order contribution in ${\bf M}_\omega$ to the vacuum interaction energy is given by the functional trace of the operator ${\bf M}_\omega$. This trace can be immediately computed from the expression for the kernel given above,
\begin{equation}
{\rm Tr}_{_{L^2}}\left({\bf M}_\omega(\alpha,\beta;{\bf r}_0)\right)=\alpha\beta\frac{G^{(0)}_\omega({\bf a}) G^{(0)}_\omega(-{\bf a})}{(1+\alpha G^{(0)}_\omega({\bf r}_0))(1+\beta G^{(0)}_\omega({\bf r}_0))}.
\end{equation}
Performing Euclidean rotation $\omega\to i\xi$ and   accounting for the expression for the free Green function in 2D, we get
\begin{equation}
{\rm Tr}_{_{L^2}}\left({\bf M}_\xi(\alpha,\beta;{\bf r}_0)\right)=\frac{2\alpha\beta K_0(a \xi ){}^2}{(\alpha  K_0(\epsilon  \xi )+2 \pi ) (\beta
   K_0(\epsilon  \xi )+2 \pi )},\label{tr-m}
\end{equation}
where $K_0(z)$ is the modified Bessel function of second kind and $0^{th}$ order. When the thickness $\epsilon$ of the quantum strings goes to zero the quantum vacuum interaction tends to zero because of the logarithmic growth, $K_0(z)\sim\log(z)$ for $z\rightarrow 0$. Therefore Dirac delta quantum strings conserve at the quantum level the property of zero interaction unlike it happens with the conical-singularity cosmic string. For higher orders in the operator ${\bf M}_\omega$ the same behaviour can be easily inferred. Hence  the Dirac delta string as mimicking a quantized cosmic string behaves as two classical cosmic strings: there is no interaction between them in the limit of zero thickness. Nevertheless as it is shown in formula (\ref{tr-m}) the quantum vacuum interaction between two Dirac delta strings is non zero when the thickness of both strings is non zero. The advantege of mimicking cosmic strings by Dirac delta potentials in 2D is that the quantum vacuum interaction between them can be expressed with a $TGTG$-formula as opposed to cosmic strings mimicked by conical singularities. The reason of this phenomenon is that the 2D Dirac delta (as well as magnetic strings) are local objects concentrated on a point that can be described as self-adjoint extensions of the Laplace operator (see reference \cite{asor13-847-852} for a modern approach) meanwhile the cosmic string described as a conical singularity is not a local object.

\section{Conclusions}
In the forgoing section we considered the vacuum polarization in the background of two parallel cosmic strings. In Section 1 we reconsidered the approach of \cite{Bordag:1990if} and identified a problem arising when dropping the 'free space' part like in other Casimir force calculations. The point is in the global character of the background leaving a dependence on the mass density of the string. As it turners out from the calculation in section 5, this contribution, which need for a regularization, can be dimensionally continued and finally evaluates to zero.

In section 2 we investigate the applicability of the 'TGTG'-formula for cosmic strings. Due to the multiplicative rather than additive way the potentials of the strings enter the equation for the Green function, this approach is not applicable here. However, we are able to evaluate the T-operator for the scattering off one string in flat coordinates, which is a non trivial task despite the simplicity of the scattering phase shift in conformal coordinates.

The vacuum interaction energy is calculated in section 5 in two ways. First, perturbatively for both strings. Here a momentum space representation can be used and a simple calculation shows a result, coinciding with earlier calculations of this kind, \cite{galt95-58-3903} and  \cite{alie97-55-3903}. The second way is to account for the first string exactly and to take only the second one perturbatively. Here the result is a quite simple formula, which was found in \cite{Bordag:1990if} with an computational mistake.

As it turns out, the disappearance of ultraviolet divergences in the vacuum interaction energy, or, equivalently, in the Casimir force, is not trivial due to the global character of the background. Also, the calculations in section 5 show divergences in intermediate steps. These were handled by introducing dimensional regularization and constructing the analytic continuation of the vacuum energy to the physical dimension. As it turned out, there is a delicate compensation of the divergences (pole terms in this case). It must be mentioned that the usual way to get a finite vacuum interaction energy (or Casimir force) by dropping a free space contribution, does not work here. For this reason we calculated the relevant heat kernel coefficients. This could be done non perturbatively using the general geometric formulas for the coefficients. As a result, there is no divergence.

In this way, we have put earlier result on the Casimir effect between cosmic strings on a more firm foundation.
\section*{Acknowledgement}
We acknowledge support from DFG in project  BO1112-18/1. We also acknowledge prof. Klaus Kirsten and prof. D. Vassilevich for valuable discussions.



\begin{thebibliography}{10}

\bibitem{Bordag:1990if}
M.~Bordag.
\newblock {On the vacuum interaction of two parallel cosmic strings}.
\newblock {\em Annalen Phys.}, 47:93, 1990.

\bibitem{line86-33-1833}
B~Linet.
\newblock {Force on a charge in the space-time of a cosmic string}.
\newblock {\em Phys.~Rev.~D}, 33:18331834, 1986.

\bibitem{lete87-4-75}
P.S. Letelier.
\newblock {Multiuple Cosmic Strings}.
\newblock {\em Class.Quant.Grav.}, 4:L75--L77, 1987.

\bibitem{dese88-118-495}
S~Deser and R~Jackiw.
\newblock {Classical and quantum scattering on a cone}.
\newblock {\em Comm.~Math.~Phys.}, 118:495--509, 1988.

\bibitem{dowk77-10-115}
J~S Dowker.
\newblock {Quantum field theory on a cone}.
\newblock {\em J.~Phys.~A: Math.~Gen.}, 10:115--124, 1977.

\bibitem{hell86-34-1918}
TM~Helliwell and DA~Konkowski.
\newblock {Vacuum fluctuations outside cosmic strings}.
\newblock {\em Phys.~Rev.~D}, 34(6):1918--1920, 1986.

\bibitem{line87-35-1987}
B~Linet.
\newblock {Quantum field theory in the space-time of a cosmic string}.
\newblock {\em Phys.~Rev.~D}, 35(2):536--539, 1987.

\bibitem{frol87-35-3779}
V.~P. Frolov and E.~M. Serebryanyi.
\newblock {Vacuum Polarization in the Gravitational Field of a Cosmic String}.
\newblock {\em Phys.~Rev.~D}, 35:3779--3782, 1987.

\bibitem{galt95-58-3903}
DV~Galtsov, YV~Grats, and AB~Lavrentev.
\newblock {Vacuum polarization and topological self-interaction of a charge in
  multiconic space}.
\newblock {\em Physics of Atomic Nuclei}, 58(3):516--521, 1995.

\bibitem{alie97-55-3903}
A.N. Aliev.
\newblock {Casimir effect in the spacetime of multiple cosmic strings}.
\newblock {\em Phys.~Rev.~D}, 55(6):3903--3906, 1997.

\bibitem{dowk89-30-770}
J~S Dowker.
\newblock {Heat Kernel Expansion On A Generalized Cone}.
\newblock {\em J.~Math.~Phys.}, 30:770, 1989.

\bibitem{furs94-11-1431}
DV~Fursaev.
\newblock {The Heat-Kernel Expansion on a Cone and Quantum-Fields Near Cosmic
  Strings}.
\newblock {\em Class.Quant.Grav.}, {11}:{1431--1443}, {1994}.

\bibitem{khus99-59-064017}
N.~R. Khusnutdinov and M.~Bordag.
\newblock {Ground state energy of massive scalar field in the background of
  finite thickness cosmic string}.
\newblock {\em Phys.~Rev.~D}, 59:064017, 1999.

\bibitem{vass03-388-279}
D.V. Vassilevich.
\newblock Heat kernel expansion: User's manual.
\newblock {\em Phys. Rept.}, 388:279--360, 2003 (arXiv:hep-th/0306138).

\bibitem{thoof88-117-685}
Gerard 't~Hooft.
\newblock {Nonperturbative Two Particle Scattering Amplitudes in
  (2+1)-Dimensional Quantum Gravity}.
\newblock {\em Comm.~Math.~Phys.}, 117:685, 1988.

\bibitem{guim94-310-297}
M.E.X. Guimaraes and B.~Linet.
\newblock {Scalar Green's functions in an Euclidean space with a conical-type
  line singularity}.
\newblock {\em Comm.~Math.~Phys.}, 310:297--310, 1994.

\bibitem{hind99-58-477}
MB~Hindmarsh and TWB Kibble.
\newblock {Cosmic strings}.
\newblock {\em Reports on Progress in Physics}, 58:477, 1999.

\bibitem{cope85-255-201}
E.J. Copeland and D.J. Toms.
\newblock Quantized antisymmetric tensor fields and selfconsistent dimensional
  reduction in higher dimensional space-times.
\newblock {\em Nucl.~Phys.~B}, 255:201--230, 1985.

\bibitem{vile85-5-263}
Alexander Vilenkin.
\newblock {Cosmic strings and domain walls}.
\newblock {\em Physics Reports}, 5(5):263--315, 1985.

\bibitem{kont38-8-1192}
M.I. Kontorovich and N.N. Lebedev.
\newblock {A method for the solution of problems in diffraction theory and
  related topics}.
\newblock {\em Zh.~Eksper.~Teor.~Fiz.}, 8(10-11):1192--1206, 1938.
\newblock (in Russian).

\bibitem{sneddon}
I.N. Sneddon.
\newblock {\em The use of integral transforms}.
\newblock McGraw Hill, New York, 1972.

\bibitem{BKMM}
M.~Bordag, G.~L. Klimchitskaya, U.~Mohideen, and V.~M. Mostepanenko.
\newblock {\em Advances in the Casimir Effect}.
\newblock Oxford University Press, 2009.

\bibitem{milton01}
K.A. Milton.
\newblock {\em {The Casimir Effect}}.
\newblock {World Scientific}, 2001.

\bibitem{full89b}
S.A. Fulling.
\newblock {\em Aspects of Quantum Field Theory in Curved Space-Time}.
\newblock Cambridge University Press, Cambridge, 1989.

\bibitem{kenn08-78-014103}
Oded Kenneth and Israel Klich.
\newblock {Casimir forces in a T-operator approach}.
\newblock {\em Phys.~Rev.~B}, {78}:{014103}, {2008}.

\bibitem{bord12-45-374012}
M.~Bordag and J.M. Munoz-Castaneda.
\newblock {Quantum vacuum interaction between two sine-Gordon kinks}.
\newblock {\em J.~Phys.~A: Math.~Gen.}, 45:374012, 2012.

\bibitem{grad94}
I.S. Gradshteyn and I.M. Ryzhik.
\newblock {\em Table of Integrals, Series and Products}.
\newblock Academic Press, New York, 1994.

\bibitem{kirs01b}
K.~Kirsten.
\newblock {\em Spectral Functions in Mathematics and Physics}.
\newblock Chapman\&Hall/CRC, Boca Raton, FL, 2001.

\bibitem{jackiw95}
R.W. Jackiw.
\newblock {\em Diverse Topics in Theoretical and Mathematical Physics, section
  I.3 (``Delta function potentials in two and three dimensional quantum
  mechanics'')}.
\newblock Advanced Series in Mathematical Physics. World Scientific, 1995.

\bibitem{godz91-59-70}
P.~Gosdzinsky and R.~Tarrach.
\newblock {Learning Quantum Field Theory From Elementary Quantum Mechanics}.
\newblock {\em Am.J.Phys.}, 59:70--74, 1991.

\bibitem{lapi82-50-45}
I.R. Lapidus.
\newblock {Quantum Mechanical Scattering In Two-Dimensions}.
\newblock {\em Am.J.Phys.}, 50:45--47, 1982.

\bibitem{Bordag:1992cm}
M.~Bordag, D.~Hennig, and D.~Robaschik.
\newblock {Vacuum energy in quantum field theory with external potentials
  concentrated on planes}.
\newblock {\em J. Phys. A}, A25:4483, 1992.

\bibitem{cast13-87-105020}
J.~M.~Munoz Castaneda, J.~Mateos~Guilarte, and A.~Moreno~Mosquera.
\newblock {Quantum vacuum energies and Casimir forces between partially
  transparent delta-function plates}.
\newblock {\em Phys.~Rev.~D}, {87}({10}):{105020}, {MAY 29} {2013}.

\bibitem{asor13-847-852}
M.~Asorey and J.~M. Munoz-Castaneda.
\newblock {Attractive and repulsive Casimir vacuum energy with general boundary
  conditions}.
\newblock {\em Nucl.~Phys.~B}, {874}({3}):{852--876}, {SEP 21} {2013}.

\end{thebibliography}
\end{document}